\documentclass[preprint,preprintnumbers,amsmath,amssymb,superscriptaddress]{revtex4}

\usepackage{graphicx}
\usepackage{dcolumn}
\usepackage{bm}

\makeatletter
\def\@seccntformat#1{}
\makeatother
\renewcommand{\numberline}[1]{}

\begin{document}
\title{A Coherent Physics Picture of Topological Insulators at Single-Particle Level}
\author{Yi-Dong Wu}
\affiliation{Department of Applied Physics, Yanshan University, Qinhuangdao, Hebei, 066004, China}
\email{wuyidong@ysu.edu.cn}

\maketitle

\textbf{The study of topological property of band insulators is an interesting branch of condensed matter physics. Two types of topologically nontrivial insulators have been extensively studied. The first type is characterized by a nonzero TKNN invariant or Chern number\cite{Thouless1982} which is directly related to the quantization of Hall conductance in the integer quantum Hall effect. Haledane proposed a model with this type of band structure in the absence of a macroscopic magnetic field\cite{Haldane1988}. We refer to such materials ``Chern insulator". The second type called ``$Z_{2}$ topological insulators" is proposed recently\cite{Kane2005a,Kane2005b}. Quantum spin Hall effect has been predicted and observed in such systems\cite{Bernevig2006,Konig2007}. Despite the intensive study there are still some fundamental problems that aren't quite clear about $Z_{2}$ insulators even at the single-particle level. For example, it's claimed that $Z_{2}$ insulators will return to its origin state after two cycles, thus coupling to the reservoirs is important for the $Z_{2}$ insulators to continuously pump spins\cite{Fu2006}. Theoretical and experimental results show quantum spin Hall effect is an edge state transport property of the materials and coupling to the reservoir seems not play an important role. So the $Z_{2}$  picture is not satisfactory in explaining these phenomena. We study the relationship between the ground state of $Z_{2}$ insulator and that of Chern insulator. Combined with the results of recent researches on polarization of Chern insulators\cite{Coh2009} and topology of edge states\cite{Fidkowski2011} we propose a coherent physics picture of topological insulators at the single-particle level.}

Unlike Chern insulators $Z_{2}$ insulators preserve the time-reversal symmetry, so the occupied state at $\mathbf{k}$ and $\mathbf{-k}$ form Kramers pairs. Time-reversal symmetry is represented by an antiunitary operator $\Theta=\exp(i\pi S_{y})K$, where $S_{y}$ is the spin operator and $K$ is complex conjunction. $\Theta$ has the property $\Theta^{2}=-1$, so there must be even number of occupied bands in $Z_{2}$ insulators. Because $Z_{2}$ insulators and Chern insulators share many similarities, it's natural to consider $Z_{2}$ insulator as a time-reversal-conserved version of Chern insulator, that is the even number of the occupied bands of ground state of $Z_{2}$ insulator can be decoupled to two ground states of Chern insulators, the two ground states are related by the time-reversal operation. In the simple case where two time-reversal related bands can be easily distinguished it's indeed the case and even the Hamiltonian can be decoupled. The two time-reversal related bands have opposite Chern numbers. According to the $Z_{2}$ classification the insulator is topological when the Chern number is odd and trivial when it's even. Unfortunately such a simple scenario can't be easily justified in general. For example, in \cite{Kane2005b} when the Rashba term is zero perpendicular spin $S_{z}$ conserves, thus two bands of up and down spins are time-reversal related and the Hamiltonian are decoupled to two Haldane's Hamiltonians. However, when Rashba term is nonzero $S_{z}$ doesn't conserve, eigenfunctions of Hamiltonian don't natually form two time-reversal related pair. So decoupling the ground states to superposition of two ground states of Chern insulators isn't obviously. In this letter we show this type of decoupling is always possible for any band insulator preserve the time-reversal symmetry.\\
To make such a decoupling we consider a two dimensional(2D) single-particle tight-binding Hamiltonian, the Hamiltonian are smoothly defined on the 2D Brillouin zone. Topologically the 2D Brillouin zone can be considered as a torus. For simplicity we assume only two bands are occupied, all our results can be easily generated to more bands. The occupied wave vectors at each $\mathbf{k}$ expand a vector space. All the occupied vector space form a vector bundle with the torus as base space. Because of time-reversal symmetry the first Chern number of the associate $U(2)$ principle bundle vanish, so the vector bundle is trivial. Two global smooth othonormal wave vectors span the vector space can be defined, we denote them $|\chi_{1}(\mathbf{k})\rangle$ and $|\chi_{2}(\mathbf{k})\rangle$. Then we construct from them two bands with the structure of ground states of Chern insulators. We show the construction of the Chern type of bands is a result of failure to construct two global continuous wave vectors satisfy  the time-reversal constraint\cite{Fu2006,Roy2009}.
\begin{eqnarray}\label{tr}
|u_{1}(\mathbf{-k})\rangle=\Theta|u_{2}(\mathbf{k})\rangle\nonumber
\\
|u_{2}(\mathbf{-k})\rangle=-\Theta|u_{1}(\mathbf{k})\rangle
\end{eqnarray}

To construct $|u_{1}(\mathbf{k})\rangle$ and $|u_{2}(\mathbf{k})\rangle$ satisfy (\ref{tr}), we focus on half of the Brillouin zone, or effective Brillouin zone(EBZ)\cite{Moore2007}. Boundaries of the EBZ are two circles corresponding to $k_{x}=0$ and  $k_{x}=\pi$. If two continuous othonormal wave vectors can be found in the EBZ and satisfy (\ref{tr}) at the boundaries, then apply the time-reversal operator on them we get two continuous othonormal wave vectors on the whole Brillouin zone. It's easy to show by an continuous $U(2)$ transformation two continuous othonormal wave vectors satisfy (\ref{tr}) can always be constructed on the two boundaries. The two continuous $U(2)$ transformation at the boundaries can't always be jointed continuously. The reason is topological. The $U(2)$ transformation can be decoupled to $U(1)\times SU(2)$ transformation, where the $U(1)$ part is the determinant of the $U(2)$ matrix. To make such a decoupling we can divide the wave vectors of one band after the $U(2)$ transformation by the determinant of the $U(2)$ matrix. Because the fundamental group of $SU(2)$ transformation $\pi_1(SU(2))=\{e\}$, the $SU(2)$ part of the transformation can be continuously jointed between the two boundaries of EBZ. The $U(1)$ part can't always be continuously jointed for they have different ``winding numbers" or more mathematically belong to different homotopy classes. When the winding number difference is even this obstruction isn't essential because the wave vectors at the boundaries satisfy (\ref{tr}) is not unique, winding number of $U(1)$ part of transformation between different choices is even, so we can make a different choice to cancel the winding number difference at the two boundaries. So globe continuous $|u_{1}(\mathbf{k})\rangle$ and $|u_{2}(\mathbf{k})\rangle$ can be defined. The odd winding number difference can't be canceled by this type of transformation at the boundary, so the insulator is inherently topological, that is it's a $Z_{2}$ insulator. It's trivial to show that $Z_{2}$ invariant is the parity of the winding number difference. We multiply the $U(1)$ part back to the origin band, say $|u_{1}(\mathbf{k})\rangle$, at the boundaries after the  continuous $SU(2)$ transformation and continuously extend the $U(1)$ transformation to the middle of EBZ. The winding number is remain constant during the extension. The U(1) transformation can be continuously defined in two separate regions of the EBZ. However they can't be continuously ``glued" together because the $U(1)$ winding number difference at the boundary of the two regions which is topologically a circle. Now we apply time-reversal operator on the $|u_{2}(\mathbf{k})\rangle$ on the EBZ, then we get the definition of $|u_{1}(\mathbf{k})\rangle$ on the other half of Brillouin zone. From (\ref{tr}) and our construction it's easy to see $|u_{1}(\mathbf{k})\rangle$ is everywhere continuous except at the boundary of two regions. By definition winding number difference is the Chern number of  $U(1)$ bundle associate with $|u_{1}(\mathbf{k})\rangle$\cite{Kohmoto1985}.  $|u_{2}(\mathbf{k})\rangle$ is defined using (\ref{tr}). Thus we finish our decoupling the ground state of $Z_{2}$ insulator to two ground states of Chern insulators. We can use the $SU(2)$ part of the transformation to  make $|u_{1}(\mathbf{k})\rangle$  and $|u_{2}(\mathbf{k})\rangle$ more smooth, eg. the first and second derivative continuous at the EBZ boundaries, so we can calculate the gauge's field and curvature on the whole Brillouin zone. There is plenty of freedom to do so. It's must be pointed out the decoupling is far from unique and even the $Z_{2}$  trivial insulators can be decoupled to two Chern insulators with even Chern numbers. The above results can be easily generalized to $2N$ bands are occupied if we notice the fundamental group $\pi_1(SU(2N))=\{e\}$.\\
With above conclusions the results of Wannier representation of the $Z_{2}$ insulators using the Kane and Mele model in \cite{Soluyanov2011} become clear. It's been shown nonzero Chern number presents a topological obstruction that prevents the construction of exponentially localized Wannier functions\cite{Brouder2007}. Our result shows if we insist two subspaces of occupied band related by the time-reversal operator the Chern number of the subspace won't be zero in the $Z_{2}$ odd case. So $Z_{2}$ is a topological obstruction to construct exponentially localized Wannier functions with time-reversal restriction.  Our results don't depend on the specific model, so they're more general.\\
Now we explain why $Z_{2}$ insulators can continuously pump spins. It's well established electric polarization is a bulk quantity can be defined in terms of Berry's Phases in ordinary insulators\cite{Vanderbilt1993}. In Chern insulators because of the present of the metallic edge states and lack of exponentially localized Wannier functions it's suspicious to use Berry's Phases to defined electric polarization. However a recent work shows the Berry-phase definition remain viable in Chern insulators and the Quantum Hall effect can be explained in terms of the adiabatic change of electric polarization under electric field\cite{Coh2009}. We use the hybrid Wannier representation in a one-band Chern insulator to illustrate the result. We choose a gauge that the band of Chern insulator is continuously defined everywhere except at $k_{x}=0(2\pi)$ where the $U(1)$ winding number difference is the Chern number $C$ of the band. Then we can define the hybrid Wannier functions
\begin{equation}
|k_{x}l_{y}\rangle=\frac{1}{2\pi}\int_{0}^{2\pi}e^{il_{y}k_{y}}dk_{y}|u(\mathbf{k})\rangle \label{hw}
\end{equation}

where $|u(\mathbf{k})\rangle$ is consistent with the gauge we choose. The centers of the hybrid Wannier functions can be expressed in term the Berry phase
\begin{equation}
y(k_{x})=\frac{1}{2\pi}\int_{0}^{2\pi}dk_{y}A_{y}(\mathbf{k})\quad (modulo\ 1) \label{hwc}
\end{equation}
where $A_{y}(\mathbf{k})=i\langle u(\mathbf{k})|\frac{\partial}{\partial k_{y}}|u(\mathbf{k})\rangle$.
By definition of Chern number $y(k_{x})$ change by the $C$ when $k_{x}$ vary from $0$ to $2\pi$. In Fig.1 we plot $y$ versus $k_{x}$.

We see the  when $k_{x}$ varies from $0$ to $2\pi$ the centers of Wannier functions form spiral instead of closed circles because of the nonzero Chern number. The total  electric polarization in y direction is the average position of center of Wannier functions multiply the charge of electron.
 \begin{equation}
P_{y}=\frac{-e}{2\pi}\int dk_{x}y(k_{x}) \quad (modulo\ -e) \label{py}
\end{equation}
 the integration about $k_{x}$ is on the whole circle and it's obviously depend on where the integration starts as is pointed out in \cite{Coh2009}. However, the adiabatic current is independent of the starting point. To get a more intuitive understanding of the adiabatic current we consider a configuration in \cite{Laughlin1981}, when the magic flux $\phi$ change by $N_x$($N_x$ is the total number of lattices along $x$ direction) flux quanta $\Delta \phi=hc/e$ the Wannier function centers will spiral one circle, so the region in Fig.1 will effectively translate $C$ lattices in $y$ direction and generate electric current. The current is obviously independent of the starting $k_{x}$. Quantum Hall effect in Chern insulator can be understood in this way. There must be gapless edge states to accommodate the pumped electrons. Since we decouple the $Z_{2}$ ground state to two Chern type of ground states of opposite Chern numbers, $Z_{2}$ insulator can be considered as two electron pumps in opposite direction. If a $Z_{2}$ insulator take the geometrical configuration in \cite{Laughlin1981}, after one pump there will be both electrons and ``holes" at the edge. The pumped electrons can't fill the holes for they are related by time-reversal operation. If $\phi$ continuously changes the two pumps will continuously pump electrons, the system certainly won't return to it's origin state after two cycles as predicted in \cite{Fu2006}. So we conclude it's the nonzero Chern numbers of the Chern insulators that the $Z_{2}$ insulators are decoupled to that pump electrons and $Z_{2}$ guarantees the Chern numbers are nonzero. In this manner band topology of $Z_{2}$ insulator guarantees the existence the time-reversal related gapless edge states. If the expectation values of spin of the edge states aren't zero the $Z_{2}$ insulator will pump spins and produce quantum spin Hall effect. \\
 A significant deficiency of $Z_{2}$ classification is some $Z_{2}$ trivial insulators can also pump spins and exhibit quantum spin Hall effect. Since $Z_{2}$ trivial insulator can be decoupled to two Chern insulators and if the insulator naturally choose a nontrivial decoupling the insulator may exhibit quantum spin Hall effect\cite{Onoda2005,Qi2006}.  For example if in one system $S_{z}$ are conserved and the spin-up and spin-down band both have nonzero even Chern numbers it will certainly pump spins despite of the fact it's $Z_{2}$ trivial\cite{Qi2006}. As we have discussed, the topology can only predict the parity of the Chern number of the two Chern bands. So band topology alone can't predict whether the gapless edge state exist or not when $Z_{2}$ is zero. The question is whether there is a natural way to decouple the ground state of a time-reversal symmetric band insulator that can determine the edge state topology. We answer the question from both the k-space point of view and the real space point of view. In k-space the Chern number is an indicator of the ``twist" of the bands.  If we want to find whether the bands are naturally twist or not we should use the bands that are ``smooth" locally, for example the wave vectors in one band can be connected by parallel transport. Natural ``twist" of bands is a geometric property of the bands. From the real space point of view, since we use the hybrid Wannier functions to calculate the polarization of Chern insulator, intuitively a more localized Wannier function will make the results more reliable. So we hope a natural decoupling should make the Wannier functions as localized as possible. The following results show the two point of view give the same way of decoupling.\\

  So far our discussions about the electron pump and the existence of the gapless edge state seem somehow more intuitive than mathematically rigorous. Fortunately a recent work on the edge states of topological insulator can be used to confirm our prediction\cite{Fidkowski2011}. In that paper topology of the edge states are identified with the topology of the spectrum of certain gluing function \begin{equation}
U_{g}(\overrightarrow{k}_{\parallel})=P\exp\left(i\int_{0}^{2\pi}A_{\perp}(\overrightarrow{k}_{\parallel},k_{\perp})dk_{\perp}\right)\label{gf}
 \end{equation}
 In our case $\overrightarrow{k}_{\parallel}$ is $k_{x}$ and $k_{\perp}$ is $k_{y}$, $A_{\perp}$ is the perpendicular(y) component of the non-Abelian Berry connection and the exponential is path ordered. The eigenfunctions of are $U_{g}(\overrightarrow{k}_{\parallel})$ are maximum localized Wannier functions for a given $\overrightarrow{k}_{\parallel}$\cite{Kivelson1982,Yu2011}.  \begin{equation}
\chi_{n}(\overrightarrow{k}_{\parallel},r-la)=\frac{1}{2\pi}\int dk_{\perp}e^{ik_{\perp}(r-la)}|u_{n}(\overrightarrow{k}_{\parallel},k_{\perp})\rangle\label{wf}
 \end{equation}
 The eigenvalues of the $U_{g}(\overrightarrow{k}_{\parallel})$ are of the form $\exp(2\pi \phi_{n})$. $\phi_{n}$, defined modulo 1, gives the real position of the center of the Wannier functions modulo the unit cell. The eigenvalues are independent the $U(2N)$ gauge. Their central claim is the edge spectrum and the spectrum ${\phi_{n}(\overrightarrow{k}_{\parallel})+l}$ have same topology. It's easy to show this Wannier function representation corresponding to a particular choice of decoupling the bands. For example in Fig 2 we plot some types of flows of $\phi_{n}$ as smooth functions of $k_{x}$ with two bands(We notice several types of flows of $\phi_{n}$ have been calculated using real models in \cite{Yu2011}). We don't limit $\phi_{n}$ to $0\leq\phi_{n}<1$, as long as it varies smoothly. The Wannier functions as the eigenfunctions of $U_{g}$ will vary smoothly except at $k_{x}=0(2\pi)$. And because of the constraint $\Theta U_{g}(\overrightarrow{k}_{\parallel}) \Theta^{-1}= U_{g}(-\overrightarrow{k}_{\parallel})^{-1}$ the pair of hybrid Wannier functions are related by the time-reversal operator. Then we transform to the k-space, clearly the two bands will be related by the time-reversal operator and they are continuous except at $k_{x}=0(2\pi)$, where there are winding number differences $\phi_{n}(2\pi)-\phi_{n}(0)$, which are the Chern numbers of the two bands.

In Fig 2, (a) (c) are corresponding to $Z_{2}$ trivial cases. In (c) the system naturally choose a Chern number 2 decoupling and there are gapless edge states. Edge states have this type of topology have been proposed in \cite{Onoda2005,Qi2006}. From topological point of view the crosses between two time-reversal symmetric points aren't protect by time-reversal symmetry. So a transformation can open ``gap" at those points. However the gapless modes are protect by the geometry of the bands. To construct the maximum localized Wannier functions the wave vectors must be parallel transported along $k_{y}$. So the maximum localized Wannier function is a geometric property of the bands. (b) (c) are corresponding to $Z_{2}$ insulator. It's easily seen no matter how to decouple the ground state as long as the Chern insulators it's decoupled to are related by time-reversal operator the center of the Wannier function won't open a ``gap". In this sense the gapless edge state of $Z_{2}$ insulator are ``strong" since it's topologically protected. Geometry of the bands determine the detail of the edge state topology in this case. In both cases we can see geometry of the bands determines the value of $Z_{2}$ and provide more information about the edges state than $Z_{2}$. Especially in the $Z_{2}$ trivial case band geometry determine whether the insulator can pump spins or not.\\

Since we have established the point of view to see the two dimensional topological insulator as two time reversal related Chern insulators and the Chern insulators can exhibits quantum Hall effect. We now propose an experiment to confirm our conclusion. Non-dissipative edge states have been observed in the HgTe quantum wells. So we consider a HgTe film of annular geometry. There is a magnetic flux $\phi$ confined to the interior of the solenoid magnet threading the hole in the annular(See Fig 3). Halperin has use the same type of geometry configuration to demonstrate the quantized Hall conductance of the quantum Hall state. The magnetic flux can pump electrons from one edge to the other. The relationship between the edge state current and the Hall potential can be established by calculating the adiabatical derivative of the total electronic energy with respect to the magnetic flux $\phi$. The difference in our case is that we propose to create a ``zero field quantum Hall state" by the adiabatically changing $\phi$. In \cite{Bernevig2006} the single-particle Hamiltonian of electrons in HgTe quantum well can be easily decoupled to two Chern insulators with up and down spin states. When $\phi$ change by one flux quanta $\Delta \phi=hc/e$ a spin up(down) electron will be pumped from the inner(outer) edge to the outer(inner) edge. First we consider the idea case when temperature is zero and the edge state is truly non-dissipative. If $\phi$ continuously change the Fermi levels of different spins will be different, the spin up(down) electrons will have a higher Fermi level at the outer(inner) edge. So there will be net currents at the edges and the currents have the same direction, that is both of them are clockwise. When magnetic flux change by $\Phi$ from the equilibrium state there will be $N=\Phi/\Delta \phi$ electrons of each spin pumped. If we consider a state not far from equilibrium the single particle density of states at the edges can be assumed to be constant $g_{in}(E_{f0})$ and $g_{ou}(E_{f0})$, which are the densities of states at the equilibrium Fermi level $E_{f0}$ at the two edges. So the Fermi levels at the edges are $E_{f0}\pm 2N/g_{in(ou)}(E_{f0})$($+(-)$ for spin up(down) electrons at inner edge and spin down(up) electron at outer edge). The total current can be calculated by
\begin{equation}
I=c \frac{\Delta U}{\Delta \phi}=4eN(1/g_{in}+1/g_{ou})/h
\end{equation}
If the magnetic flux $\Phi$ hold constant there will be constant superconducting currents on the edges. We call this state a ``zero field quantum Hall state". It's actually can be viewed as two quantum Hall states put together. First instead of external magnetic field the spin-obit coupling generate the nonzero TKNN or Chern number for the two time-reversal related Chern states. Second there are no net charges at the edges, so there are no macro electrostatic field in the film or we can consider it as the two quantum Hall states have opposite electrostatic fields and when they put together the electric fields cancel. For the two quantum Hall states have opposite Chern number and opposite electric fields it's easy to anticipate the currents are in the same direction. This state is a direct evidence of the quantum spin Hall effect. Unlike the previous experiment, in which only electric transport properties of the edge states are measured, the existence of the edge currents in our case means there are net spins on the two edges and the topological insulators can pump spins. Because the film aren't connected to any contacts or reservoirs it shows the $Z_2$ picture is inaccurate, which assert topological insulators can't pump spins unless connected to reservoirs. Though we construct the experiment in the HgTe quantum well in which $S_z$ is conserved. From our general theory the ``zero field quantum Hall state" can be established in any two dimensional topological insulators.\\

In conclusion, we propose a clear physics picture of $Z_{2}$ insulator at single-particle level that is to view the $Z_{2}$ insulator as two time-reversal related Chern insulators. The two Chern insulators can pump electrons to opposite directions and produce quantum spin Hall effect. The picture is discussed both intuitively and with mathematical rigor. We find the gapless edge state in some $Z_{2}$ trivial insulators are protected by band geometry. So the quantum spin Hall effect can occur in such materials.

\begin{figure}[h]
\includegraphics[width=5 in,clip=true]{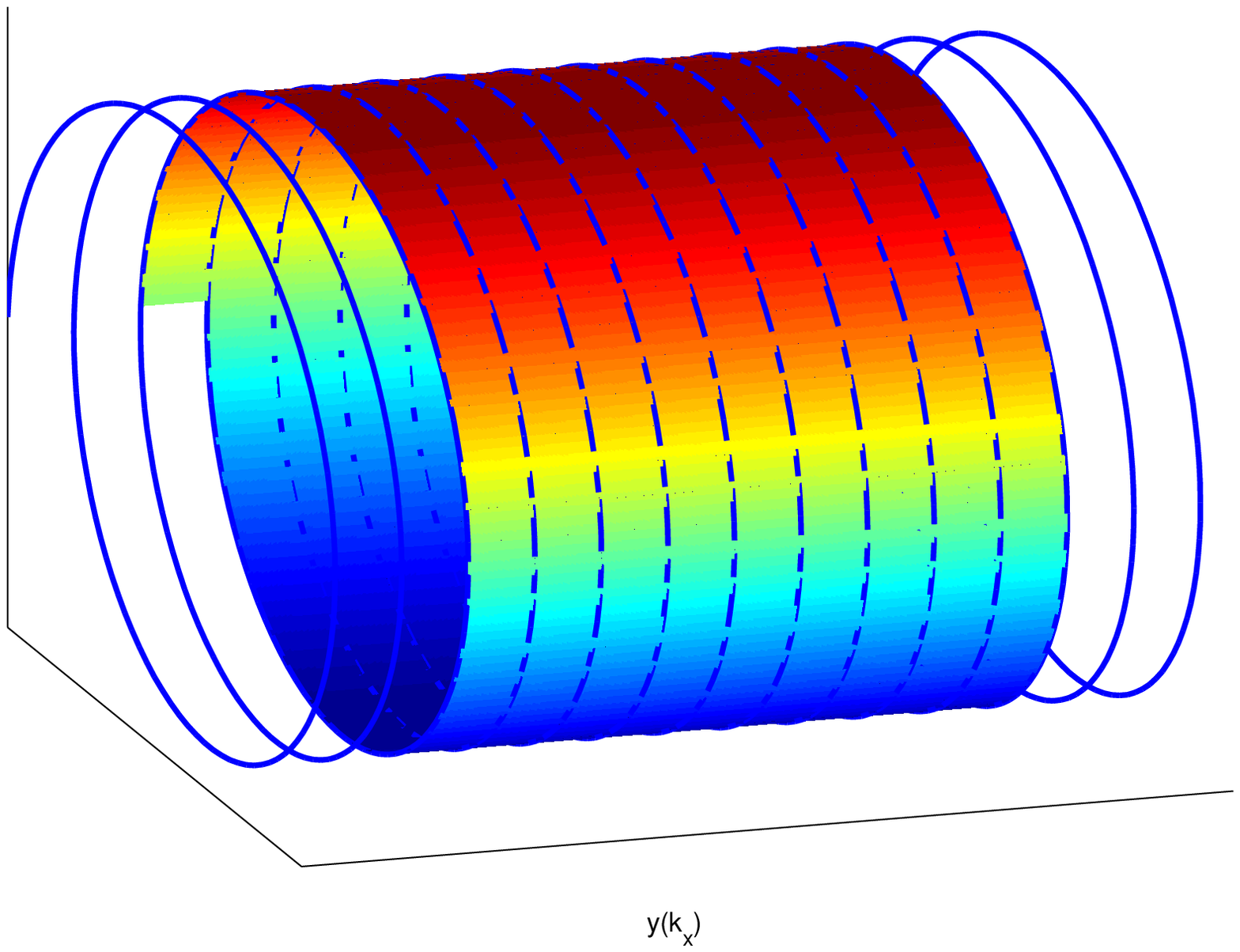}
\caption{Position of the hybrid Wannier function center $y$ as function of $k_{x}$. $k_{x}$ is drawn as the angular coordinate in the plane perpendicular to y. Chern number $C=1$.} \label{wac}
\end{figure}

 \begin{figure}[h]
\includegraphics[width=5 in,clip=true]{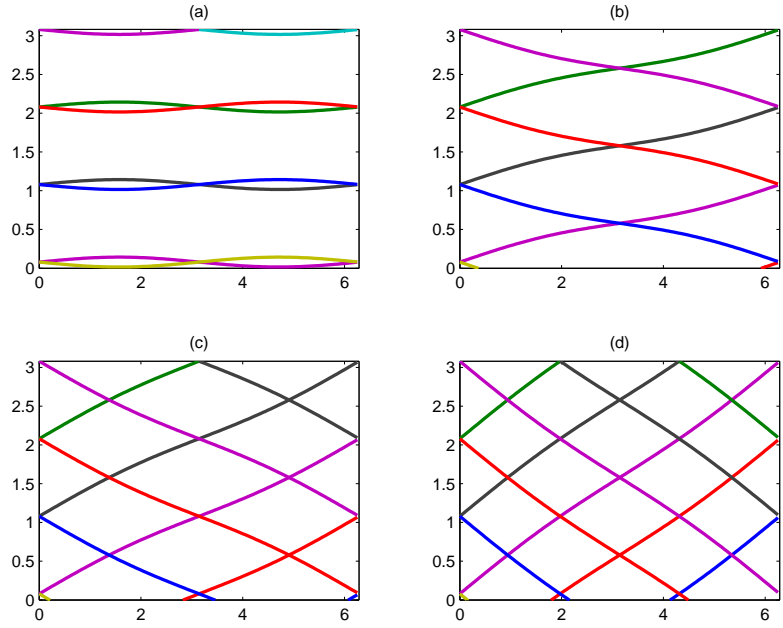}
\caption{Four types of flows of $\phi_{n}$ with two bands. The corresponding Chern number of the bands are $\pm 0-3$} \label{phi}
\end{figure}
\begin{figure}[h]
\includegraphics[width=5 in,clip=true]{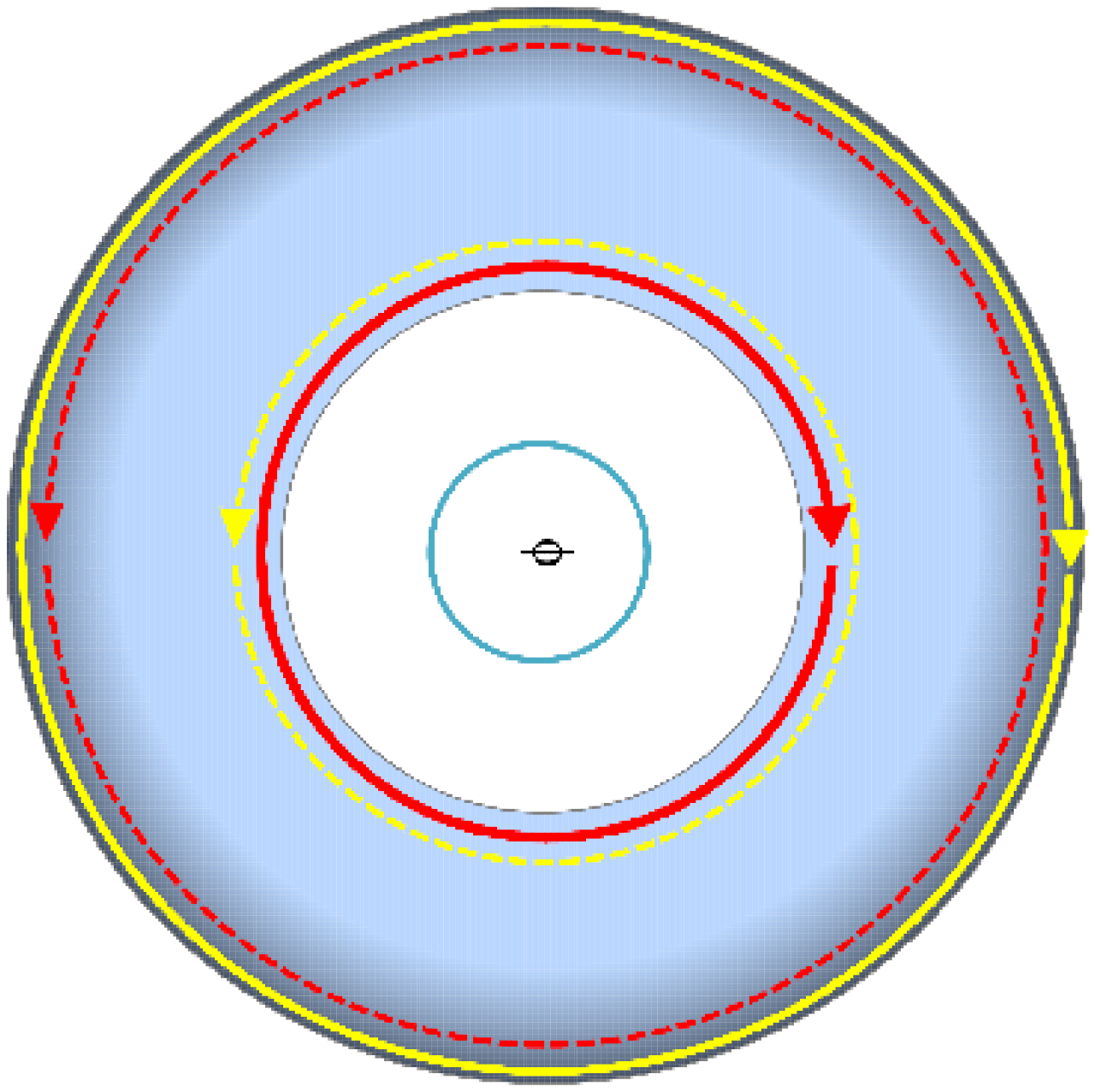}
\end{figure}

\end{document}